\documentclass{article}
\usepackage{amsmath}
\usepackage{cite}
\def\Ref#1{(\ref{#1})}
\begin{document}
\begin{titlepage}

\hfill{}
\vskip 5 mm
\noindent{ \Large \bf Phase transition in an asymmetric generalization of
           the zero-temperature $q$-state Potts  model}

\vskip 1 cm \noindent{N. Majd$^1$, A. Aghamohammadi$^{1,a}$, \& M.
Khorrami$^{2,b}$} \vskip 5 mm {\it
  \noindent{ $^1$ Department of Physics, Alzahra University,
             Tehran 19834, Iran. }

  \noindent{ $^2$ Institute for Advanced Studies in Basic Sciences,
             P.O.Box 159, Zanjan 45195, Iran. }

  \noindent{ $^a$ mohamadi@theory.ipm.ac.ir}

  \noindent{ $^b$ mamwad@iasbs.ac.ir}  }
\vskip 1 cm



\vskip 1cm

\begin{abstract}
An asymmetric generalization of the zero-temperature $q$-state
Potts model  on a one dimensional lattice, with and without
boundaries, has been studied. The dynamics of the particle number,
and specially the large time behavior of the system has been
analyzed. In the thermodynamic limit, the system exhibits two
kinds of phase transitions, a static and a dynamic phase
transition.

\end{abstract}
\end{titlepage}
%%%%%%%%%%%%%%%%%  The body of the paper            %%%%%%%%%%%%%%%%%%%%%
\section{Introduction}
In recent years, reaction-diffusion systems have been studied by
many people. As mean-field techniques, generally, do not give
correct results for low-dimensional systems, people are motivated
to study exactly-solvable stochastic models in low dimensions.
Moreover, solving one-dimensional systems should in principle be
easier. Exact results for some models on a one-dimensional lattice
have been obtained, for example in
\cite{ADHR,KPWH,HS,PCG,HOS1,HOS2,AL,AKK,RK,RK2,AKK2,AAMS}, and in
\cite{RK3,SAM} for the case of multi-species systems. Different
methods have been used to study these models, including analytical
and asymptotic methods, mean field methods, and large-scale
numerical methods.

Some interesting problems in non-equilibrium systems are
non-equilibrium phase transitions described by phenomenological
rate equations, and the way the system relaxes to its steady
state. Kinetic generalizations of the Ising model, for example the
Glauber model or the Kawasaki model, are such phenomenological
models and have been studied extensively
\cite{RG,KK,DHP,DHP2,BFK}. Combination of the Glauber and the
Kawasaki dynamics has been also considered \cite{AF,TV,SSG}. One
of the models which has been extensively studied is the $q$-state
Potts model evolving according to some generalization of the
zero-temperature Glauber dynamics (see \cite{GL,DRS,GF,AT,GS,MbA},
for example). The evolution tends to align all the spins, and so
domains of parallel spins grow with time.

In this paper, we want to study an asymmetric generalization of
the zero-temperature $q$-state Potts  model on an infinite lattice
with and without boundaries. There are also reactions at the
boundaries. The dynamics of the particle number, and specially the
large time behavior of the system is studied. In the thermodynamic
limit, the system shows two kinds of phase transitions. One of
these is a static phase transition, the other a dynamic one. The
static phase transition is controlled by the reaction rates, and
is a discontinuous change of the behavior of the derivative of the
stationary particle density at the end points, with respect to the
reaction rates. The dynamic phase transition is controlled by the
reaction rates at boundaries, and is a discontinuous change of the
relaxation time towards the stationary configuration. There is a
\textit{fast phase} and a \textit{slow phase}. Increasing the
reaction rates at the boundaries, the system goes from the
\textit{slow phase} to the \textit{fast phase}. This model may be
considered as a biased voting model, in the sense that there are
$q$ different opinions, and when people at the adjacent sites have
different opinions, they may interact so that their opinions
become the same.
\section{Asymmetric $q$-state Potts model at zero temperature}
In the ordinary Glauber model, the interaction is between three
neighboring sites. Spin flip brings the system to equilibrium with
a heat bath at temperature $T$. A spin is flipped with the rate
$\mu:=1-\tanh[J/(kT)]$, if the spin of both of its neighboring
sites are the same as itself; and is flipped with the rate
$\lambda:=1+\tanh[J/(kT)]$, if the spin of both of its neighboring
sites are opposite to it. At domain boundaries, the spins are
flipped with unit rate. So the interactions can be written as,
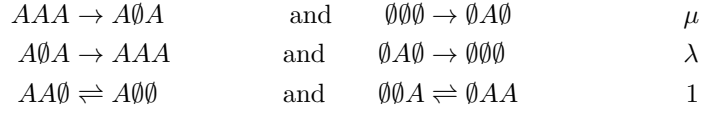
\begin{align}
AAA&\to A\emptyset A &\hbox{and}\qquad \emptyset\emptyset\emptyset
&\to \emptyset A\emptyset\qquad &\mu\nonumber\\
A\emptyset A &\to AAA &\hbox{and}\qquad \emptyset A\emptyset &\to
\emptyset\emptyset\emptyset\qquad &\lambda\nonumber\\
AA\emptyset &\rightleftharpoons A\emptyset\emptyset
&\hbox{and}\qquad\emptyset\emptyset A &\rightleftharpoons
\emptyset AA\qquad &1\nonumber
\end{align}
where spin up and spin down are denoted by $A$ and $\emptyset$.
One can interpret an up spin as a  particle, and a down spin as
a hole. At zero temperature, the Glauber dynamics is effectively a
two--site interaction \cite{GS}:
\begin{equation}
A\emptyset\to (AA,\;\emptyset\emptyset)\qquad \emptyset A\to (AA,
\;\emptyset\emptyset)
\end{equation}
where all the above processes occur with  the same rate.

One can consider the following interactions, as an asymmetric
generalization of the zero-temperature Glauber model.
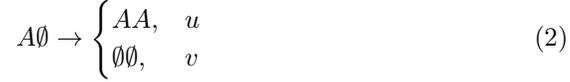
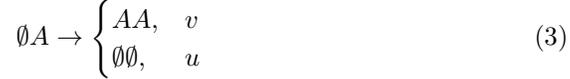
\begin{equation}
A\emptyset\to
\begin{cases}
AA, &u\\
\emptyset\emptyset, &v
\end{cases}
\end{equation}
and
\begin{equation}
\emptyset  A  \to
\begin{cases}
AA, &v\\
\emptyset\emptyset, &u
\end{cases}
\end{equation}
If $u\ne v$, the above system has left-right asymmetry. The above
system on an infinite lattice has been investigated in \cite{AM},
where its $n$-point functions, its equilibrium states, and its
relaxation towards these states are studied. It can be easily
shown that the time evolution equation for the average
particle-numbers of the system with the above interactions are the
same as that of a system with the following interactions, where
diffusion is also present:
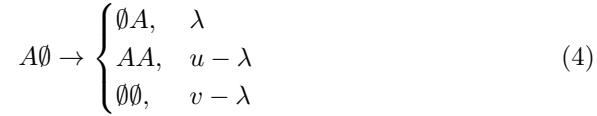
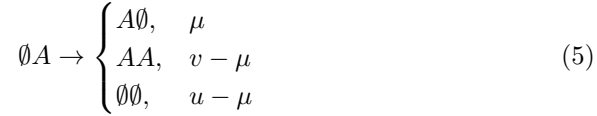
\begin{equation}\label{e1}
A\emptyset\to
\begin{cases}
\emptyset A, &\lambda\\
AA, &u-\lambda\\
\emptyset\emptyset, &v-\lambda
\end{cases}
\end{equation}
and
\begin{equation}\label{e2}
\emptyset A\to
\begin{cases}
A\emptyset, &\mu\\
AA, &v-\mu\\
\emptyset\emptyset, &u-\mu
\end{cases}
\end{equation}

The model addressed in this article is an asymmetric
generalization of the zero-temperature $q$-state Potts model in
one dimension and with Glauber dynamics. The interaction is a
nearest neighbor interaction and is defined as follows
\begin{equation}\label{1}
A_\alpha A_\beta\to
\begin{cases}
A_\alpha A_\alpha, &u\\
A_\beta A_\beta, &v
\end{cases}
\end{equation}
Similar to \Ref{e1} and \Ref{e2}, adding diffusion do not alter
the time evolution equation of the average particle numbers. The
symmetric case ($u=v$) has been recently analyzed using a
generalization of the empty interval method \cite{MbA}. The
one-species case ($q=2$) has been studied in \cite{AM} (without
boundaries) and \cite{MA} (with boundaries). It is shown that for
the infinite lattice without boundaries, the system has two ground
state; The sites are all occupied or all vacant. This is
reminiscent of the ordinary Glauber model, and asymmetry does not
change the ground states. Denoting by $|0\rangle$
($|\Omega\rangle$) the empty (full) lattice, the state of the
system at infinitely large times is
\begin{equation}\label{2}
|P(\infty )\rangle =(1-\rho_0 )|0\rangle +\rho_0 |\Omega\rangle.
\end{equation}
where $\rho_0$ is the initial average density. For the asymmetric
generalization of the $q$-state Potts model, there are $q$ ground
states. To see this, one can divide the state at each site to two
classes, $A_1$ and not-$A_1$, for example. The state not-$A_1$ is
denoted by $B_1$. So the interactions are
\begin{equation}\label{3}
A_1 B_1\to
\begin{cases}
A_1 A_1,&u\\
B_1 B_1,&v
\end{cases}
\end{equation}
and
\begin{equation}
B_1 A_1\to
\begin{cases}
B_1 B_1,&u\\
A_1 A_1,&v
\end{cases}
\end{equation}
Similar to the asymmetric zero-temperature Glauber model, the
ground states are the states that all the sites of the lattice are
in the state $A_1$, or none of the sites are in the state $A_1$.
Repeating the same argument for the substates of $B_1$, it is seen
that there are $q$ ground states. Each ground state is a state in
which the states of all lattice sites are the same state,
$A_\alpha$ for example. So the final state is
\begin{equation}\label{4}
|P(\infty)\rangle =\sum_\alpha\rho_0^\alpha |A_\alpha\rangle,
\end{equation}
where $|A_\alpha\rangle$ is the state for which all the sites are
in the state $A_\alpha$, and $\rho_0^\alpha$ is the initial
average density of the state $A_\alpha $. Then, all the
correlation functions at infinitely large times can be obtained
easily:
\begin{equation}\label{5}
\langle n_i^\alpha n_j^\beta\cdots
n_k^\gamma\rangle=\delta^{\alpha\beta}\delta^{\alpha\gamma}\cdots\rho_0^\alpha.
\end{equation}

The evolution equation for the average number of the state
$\alpha$ at the site $j$, $\langle n_j^\alpha \rangle$ is
\begin{equation} \label{6}
\langle \dot n_j^\alpha \rangle = -(u+v)\langle  n_j^\alpha\rangle
+u \langle n_{j-1}^\alpha\rangle + v\langle  n_{j+1}^\alpha\rangle
\end{equation}
The above equation shows that the evolution equations for the
average number of different states are decoupled, and so the
average number of each state depends only on the initial value of
the average number of that state. In fact,
\begin{equation}\label{7}
\langle n_j^\alpha (t) \rangle =e^{-(u+v)t}\sum_m\left({u\over
v}\right)^{m-j\over 2}I_{m-j}(2\sqrt{uv} t)\langle  n_j^\alpha (0)
\rangle,
\end{equation}
and at large times,
\begin{equation}\label{8}
\langle n_j^\alpha (t) \rangle-\rho_0^\alpha\sim\left({u\over
v}\right)^{j\over 2}{e^{[-(u+v)+2\sqrt{uv}]t}\over \sqrt{t}}.
\end{equation}
It is seen from the above equation that if $u<v$, the expectation
at the rightmost sites tend rapidly to their final value, and
obviously for $u>v$ the leftmost sites arrive earlier to their
final states.
\section{$q$-state Potts model on a lattice with the boundaries}
In this section, we study $q$-state Potts model on a lattice with
reaction at the boundaries. The interactions on the bulk of
lattice are \Ref{1}. The exchange of the states  at the first site
is
\begin{equation}\label{9}
  A_\beta \to A_\alpha \qquad \hbox{with the rate
  $\Lambda^\alpha_\beta$},
\end{equation}
and at the final site, it is
\begin{equation}\label{10}
  A_\beta \to A_\alpha \qquad \hbox{with the rate $\Gamma^\alpha_\beta$}.
\end{equation}
For $ \alpha \ne \beta$, $\Lambda^\alpha_\beta$ and
$\Gamma^\alpha_\beta$ are rates and should be nonnegative. The
diagonal elements of $\Lambda$ and $\Gamma$ are chosen so that
\begin{equation}\label{11}
s_\alpha\Lambda^\alpha_\beta=s_\alpha\Gamma^\alpha_\beta=0,
\end{equation}
where
\begin{equation}\label{ha}
s_\alpha=1.
\end{equation}
From now on, a repeated subscript and superscript imply a
summation over the repeated index.

The equations of motion for the average numbers are
\begin{align}\label{12}
\langle \dot n_j^\alpha \rangle &= -(u+v)\langle n_j^\alpha\rangle
+u \langle n_{j-1}^\alpha\rangle + v\langle
n_{j+1}^\alpha\rangle,\quad j\ne 1,L\nonumber\\
\langle \dot n_1^\alpha \rangle &=\Lambda^\alpha_\beta \langle
n_1^\beta\rangle -v \langle n_{1}^\alpha\rangle + v\langle
n_{2}^\alpha\rangle,\nonumber \\
\langle \dot n_L^\alpha \rangle &=\Gamma^\alpha_\beta \langle
n_L^\beta\rangle -u \langle n_{L}^\alpha\rangle + u\langle
n_{L-1}^\alpha\rangle.
\end{align}
Defining the vector $N_k$ through
\begin{equation}\label{13}
N_k^\alpha:=\langle n_k^\alpha\rangle
\end{equation}
equation \Ref{12} takes the form
\begin{align}\label{14}
\dot N_k &= -(u+v)N_k +u N_{k-1}+ v
N_{k+1},\quad k\ne 1,L\nonumber \\
\dot N_1 &= {\bf\Lambda}N_1-v N_{1}+ v
N_{2},\nonumber \\
\dot N_L&= {\bf\Gamma}N_L-u N_{L}+ uN_{L-1}.
\end{align}
For the stationary state,
\begin{equation}\label{15}
\dot N_k=0.
\end{equation}
Substituting the ansatz
\begin{equation}\label{16}
N_k=B z_1^k+C z_2^k
\end{equation}
in the first equation of \Ref{14}, one arrives at
\begin{equation}\label{17}
-(u+v)+vz_i+{u\over{z_i}}=0,
\end{equation}
whose solutions are $z_1=1,\quad z_2=u/v$. First assume $u<v$.
Then, one can write \Ref{16} as
\begin{equation}\label{18}
N_k=B+C'\left({u\over v}\right)^{k-1}.
\end{equation}
Since $s_\alpha N_k^\alpha =1$, it is seen that, $B$, and $C'$
satisfy
\begin{align}\label{19}
s_\alpha B^\alpha&=1,\nonumber\\
s_\alpha {C'}^\alpha&=0.
\end{align}
Using the second and third equations of \Ref{14}, one arrives at
\begin{align}\label{20}
(\Lambda-v)\left[B+C'\right]+v\left[B+C'
\left({u\over v}\right)\right]&=0,\nonumber\\
(\Gamma-u)\left[B+C'\left({u\over
v}\right)^{L-1}\right]+u\left[B+C'\left({u\over
v}\right)^{L-2}\right]&=0.
\end{align}
In the thermodynamic limit ($L\to \infty$), these two equations
yield
\begin{align}\label{21}
\Gamma B&=0,\nonumber\\
(\Lambda-v+u)C'&=-\Lambda B.
\end{align}
The first equation in \Ref{21} can be used to determine $B$.
Substituting it in the second one, $C'$ is obtained. The
degeneracy of the zero eigenvalue of $\Gamma$ is equal to the
degeneracy of the stationary average particle number. The matrix
$\Gamma$ does have a zero eigenvalue, since $s$ is the left
eigenvector of $\Gamma$ with the eigenvalue zero. If the zero
eigenvalue of $\Gamma$ is degenerate, then the final value of the
average particle numbers depend on the initial conditions. Also
note that the second equation of \Ref{21} has one and only one
solution for $C'$. The reason is that the real part of the
eigenvalues of $\Lambda$ are nonpositive. So, $v-u$ is not an
eigenvalue of $\Lambda$, and hence the matrix $\Lambda-v+u$ is
nonsingular.

If $u>v$, then the term $(u/v)^k$ diverges for the right end
sites. So, we choose
\begin{equation}\label{18b}
N_k=B+C''\left({u\over v}\right)^{k-L}.
\end{equation}
In the thermodynamics limit, one arrives at
\begin{align}\label{21a}
\Lambda B&=0,\nonumber\\
(\Gamma+v-u)C''&=-\Gamma B.
\end{align}
If $u>v$, it is the degeneracy of the zero eigenvalue of
$\Lambda$, which determines the degeneracy of the stationary
states of the particle numbers. Here too, the second equation of
\Ref{21a} determines $C''$ uniquely.

If $u<v$, the profile of average particle number is flat for the
rightmost sites. As $u$ exceeds $v$, it acquires a finite slope,
proportional to $\ln (u/v)$. For the leftmost sites, this behavior
is reversed. In fact, there is a discontinuous change of the
behavior of the derivative of the stationary particle number at
the end points. This is a static phase transition, and it is
controlled only by the reaction rates and does not depend on the
reaction rates at the boundaries. The case $q=2$ has been already
studied in \cite{MA}. One notes that the matrices $\Lambda$ and
$\gamma$ are
\begin{align}
\Lambda&=
  \begin{pmatrix}
    -a' & a \\
    a' & -a
  \end{pmatrix},
\nonumber\\
\Gamma&=
  \begin{pmatrix}
    -b' & b \\
    b' & -b
  \end{pmatrix},
\end{align}
in terms of the appropriate parameters of that references. Here,
the zero eigenvalues of $\Lambda$ and $\Gamma$ are nondegenerate
(unless the rates are zero).

Now let's consider the relaxation of the system towards its
stationary state. The homogeneous part of \Ref{12} is
\begin{equation}\label{22}
\dot\delta N_k^\alpha=h_{k\beta}^{l\alpha}\delta N_l^\beta,
\end{equation}
where $\delta N$ is the difference of $N$ from the its stationary
value, so that
\begin{equation}\label{del}
s_\alpha\delta N_k^\alpha=0.
\end{equation}
To study the relaxation of the system, we investigate the
eigenvalues of $h$. One finds
\begin{align}\label{23}
E x_k^\alpha =&-(u+v)x_k^\alpha +u x_{k-1}^\alpha +v
x_{k+1}^\alpha ,\qquad k\ne 1,L\nonumber \\
E x_1^\alpha =& \Lambda^\alpha_\beta x_1^\beta -vx_1^\alpha+v
x_2^\alpha,\nonumber \\
E x_L^\alpha =& \Gamma^\alpha_\beta x_L^\beta -u x_L^\alpha +u
x_{L-1}^\alpha ,
\end{align}
where the eigenvalue and eigenvector have been denoted by $E$ and
$x$, respectively. The solution to these equations is
\begin{equation}\label{24}
x_k^\alpha ={\cal B}^\alpha z_1^k+{\cal C}^\alpha z_2^k,
\end{equation}
where $z_i$'s satisfy
\begin{equation}\label{25}
E=-(u+v)+vz_i+{u\over{z_i}}.
\end{equation}
Performing the change of variable $z_i=:\sqrt{u/v}Z_i$, the second
and third equations of \Ref{23} take the form
\begin{align}\label{26}
\sqrt{u\over v}E({\cal B}Z+{\cal C}Z^{-1})=&( \Lambda
-v)\sqrt{u\over v}({\cal B}Z+{\cal C}Z^{-1})+v\left(\sqrt{u\over v
}\right)^2({\cal B}Z^2+{\cal C}Z^{-2})\nonumber
\\
\sqrt{u\over v}E({\cal B}Z^L+{\cal C}Z^{-L})=&( \Gamma
-u)\left(\sqrt{u\over v}\right)^L({\cal B}Z^L+{\cal
C}Z^{-L})\nonumber\\
&+u\left(\sqrt{u\over v}\right)^{L-1}({\cal B}Z^{L-1}+{\cal
C}Z^{-L+1})
\end{align}
Substituting $E=-(u+v)+\sqrt{uv}(Z+Z^{-1})$ in the above
equations, one can write them in the matrix form
\begin{equation}\label{27}
  \begin{pmatrix}
    -(u+\Lambda )Z+\sqrt{uv} & -(u+\Lambda )Z^{-1}+\sqrt{uv} \\
                             &\\
    -(v+\Gamma )Z^L+\sqrt{uv}Z^{L+1} & -(v+\Gamma )Z^{-L}+\sqrt{uv}Z^{-L-1}
  \end{pmatrix}
  \begin{pmatrix}
    {\cal B} \\
    \\
    {\cal C}
  \end{pmatrix}
=0.
\end{equation}
To have nontrivial solutions for ${\cal B}$ and ${\cal C}$, the
determinant of the matrix should be zero:
\begin{equation}\label{a38}
\det\begin{pmatrix}
    -(u+\Lambda )Z+\sqrt{uv} & -(u+\Lambda )Z^{-1}+\sqrt{uv} \\
                             &\\
    -(v+\Gamma )Z^L+\sqrt{uv}Z^{L+1} & -(v+\Gamma )Z^{-L}+\sqrt{uv}Z^{-L-1}
  \end{pmatrix}
=0.
\end{equation}
In the thermodynamic limit ($L\to\infty$), if all of the roots of
\Ref{a38} are phases, the largest real part of the eigenvalues of
the Hamiltonian will be $-(u+v)+2\sqrt{uv}$, from which the
relaxation time of the system is
\begin{equation}\label{a39}
\tau_{\rm f}=(u+v-2\sqrt{uv})^{-1}.
\end{equation}
We call this the \textit{fast phase}.

If some of the roots of \Ref{a39} are not phases, however, this
may be not the case: there may be an eigenvalue of the Hamiltonian
with a larger real part, which corresponds to a larger relaxation
time. For $Z=r\; e^{i\theta}$, one has
\begin{equation}\label{a40}
\Re(E)=-(u+v)+\sqrt{uv}\left(r+{1\over r}\right)\cos\theta.
\end{equation}
This leads to a larger relaxation time, provided
\begin{equation}\label{a41}
\left(r+{1\over r}\right)\cos\theta>2.
\end{equation}
If there is such a solution for $Z$, the system is said to be in
the \textit{slow phase}.

So, if none of the roots of \Ref{a38} satisfy \Ref{a41}, then the
system is in the fast phase, and the relaxation time does not
depend on the reaction matrices $\Gamma$ and $\Lambda$. Otherwise,
the relaxation time does depend on the reaction matrices (the slow
phase). The transition between these two phases is the dynamical
phase transition.

Now, let's seek the nonphase solutions of \Ref{a38}. If $Z$ is a
solution to \Ref{a38}, $1/Z$ is another solution to it. So it is
sufficient to seek the solutions with $|Z|>1$. In the
thermodynamic limit, and for $|Z|>1$, \Ref{a38} is simplified to
\begin{equation}\label{28}
  \det [\sqrt{uv}Z-(v+\Gamma )]\det[\sqrt{uv}Z-(u+\Lambda )]=0,
\end{equation}
which consists of the characteristic equations for $\Gamma$ and
$\Lambda$. Denoting the eigenvalues of $\Gamma$ and $\Lambda$ by
$\gamma$ and $\lambda$, respectively, one has
\begin{equation}\label{29}
Z={v+\gamma\over \sqrt{uv}}, \quad {\rm or} \quad
Z={u+\lambda\over\sqrt{uv}}.
\end{equation}
As $\Gamma$ and $\Lambda$ are stochastic matrices, their
eigenvalues have nonpositive real-parts. For $u\geq v$, the real
part of $(v+\gamma)/\sqrt{uv}$ is then not greater than 1. Hence
it cannot satisfy \Ref{a41}. So, the only relevant equation for
finding the system in the \textit{slow phase} is
\begin{equation}\label{a42}
Z={{u+\lambda}\over{\sqrt{uv}}}.
\end{equation}
A similar argument shows that for $v\geq u$, the first equation of
\Ref{29} is relevant. One also concludes that for $u=v$, the
system has no \textit{slow phase}.

So, without loss of generality, let's take $u<v$. As $\Gamma$ is a
stochastic matrix, it has at least one zero eigenvalue. However,
if this eigenvalue is nondegenerate, the right eigenvector of
$\Gamma$ ($\cal B$) cannot satisfy
\begin{equation}\label{a43}
s_\alpha{\cal B}^\alpha=0,
\end{equation}
since $s$ is the left eigenvector of $\Gamma$ corresponding to the
same eigenvalue. But from \Ref{del}, it is seen that $\cal B$
should satisfy \Ref{a43}. So, from the eigenvalues of $\Gamma$,
one should set aside one zero eigenvalue, and consider only the
other eigenvalues.

The system undergoes a dynamic phase-transition at the point that
for one of the $Z$'s in \Ref{a42} the criterion \Ref{a41} is
satisfied. At this point, the real- and imaginary-parts of $Z$
satisfy
\begin{equation}\label{47}
Y=\pm(X-1)\sqrt{X\over{2-X}},\qquad X>1,
\end{equation}
where
\begin{align}\label{48}
X&:=\Re(Z),\nonumber\\
Y&:=\Im(Z).
\end{align}
One can translate this in terms of $\lambda$. The criterion for
the \textit{slow phase} is then seen to be
\begin{equation}\label{49}
|\Im(\lambda)|<\left[\Re(\lambda)+u-\sqrt{uv}\right]
\sqrt{{\Re(\lambda)+u}\over{2\sqrt{uv}-\Re(\lambda)-u}},\hbox{
or}\quad\Re(\lambda)>2\sqrt{uv}-u.
\end{equation}

A simple way to induce the phase transition is to multiply the
matrix $\Lambda$ by a parameter $r$. This means multiplying the
rates of the reaction at the first site by $r$. As
$\Re(\lambda)\leq 0$, one can see that for a large enough value of
$r$, the value of $\Re(\lambda)+u-\sqrt{uv}$ will be negative
(provided $\Re(\lambda)\ne 0$, that is, provided the zero
eigenvalue of the matrix $\Lambda$ is not degenerate). So the
system will be in the \textit{fast phase}. It is also seen that as
$r$ tends to zero, either $2\sqrt{uv}-\Re(\lambda)-u$ becomes
negative, or in the first inequality in \Ref{49} the right-hand
becomes greater than the left-hand side (which tends to zero). So,
the system will be in the \textit{slow phase}. Roughly speaking,
increasing the reaction rates brings the system from the
\textit{slow phase} to the \textit{fast phase}. A similar argument
holds, of course, for the case $v>u$ and the eigenvalues of the
matrix $\Gamma$.

\vskip 2\baselineskip
\noindent{\bf Acknowledgement}\\
A. A. and M. K. would like to thank Institute for Studies in
Theoretical Physics and Mathematics for partial support.
\newpage
%%%%%%%%%%%%%%%%%%%%%%%%%%%%%%%%%%

\end{document}